\documentclass[12pt,preprint]{aastex}

\usepackage{natbib}
\usepackage{rotating}
\begin{document}

\title{Detection of a Molecular Disk Orbiting the Nearby, ``Old,''
  Classical T Tauri Star MP Mus\footnote{This research is 
  based on observations collected at
  the European Organisation for Astronomical Research in the Southern
  Hemisphere, Chile, proposal number 385.C-0143, with the Atacama
  Pathfinder Experiment (APEX).}}

\author{Joel H. Kastner\altaffilmark{1}, Pierry Hily-Blant\altaffilmark{2},  
  G. G. Sacco\altaffilmark{1}, Thierry
  Forveille\altaffilmark{2}, B. Zuckerman\altaffilmark{3}}

\altaffiltext{1}{Center for Imaging Science, Rochester Institute of
  Technology, 54 Lomb Memorial Drive, Rochester NY 14623
  (jhk@cis.rit.edu)} 
\altaffiltext{2}{Laboratoire d'Astrophysique de
  Grenoble, Universit\'e Joseph Fourier --- CNRS, BP 53, 38041
  Grenoble Cedex, France} 
\altaffiltext{3}{Dept.\ of Physics \& Astronomy,
  University of California, Los Angeles 90095, USA}

\begin{abstract}
  We have used the Atacama Pathfinder Experiment 12 m telescope to
  detect circumstellar CO emission from MP Mus (K1 IVe), a nearby
  ($D\sim100$ pc), actively accreting, $\sim7$ Myr-old pre-main
  sequence (pre-MS) star. The CO emission line profile measured for MP Mus is
  indicative of an orbiting disk with radius $\sim120$ AU,
  assuming the central star mass is $1.2$ $M_\odot$ and the disk
  inclination is $i \sim 30^\circ$, and the inferred disk molecular
  gas mass is $\sim3 M_\earth$. MP Mus thereby joins TW Hya and V4046
  Sgr as the only late-type (low-mass), pre-MS star systems within
  $\sim$100 pc of Earth that are known to retain orbiting, molecular
  disks.  We also report the nondetection (with
  the Institut de Radio Astronomie Millimetrique 30 m telescope) of CO
  emission from another ten nearby ($D\stackrel{<}{\sim}100$ pc),
  dusty, young (age $\sim10-100$ Myr) field stars of spectral type
  A--G. We discuss the implications of these results for the
  timescales for stellar and Jovian planet accretion from, and
  dissipation of, molecular disks around young stars.
\end{abstract}

\section{Introduction}

Over the past two decades, astronomers have identified a few hundred
young (ages $\sim5$ to $\sim$100 Myr) stars, including more than a
half-dozen T Tauri and post-T Tauri associations, within $\sim$100 pc
of Earth \citep[][and references
therein]{1997Sci...277...67K,2004ARA&A..42..685Z,2008hsf2.book..757T}. Most
such stars have been identified on the basis of their
characteristically large X-ray fluxes, thermal infrared emission from
warm circumstellar dust, optical spectral features (principally,
strong Li absorption lines), and/or distinctive space velocities
\citep[see discussion in][]{2004ARA&A..42..685Z}.
Based on the low frequency of mid- to far-infrared excesses among the
nearby, young stellar groups, only a small percentage of such stars
evidently retain massive, dusty, circumstellar disks
\citep[e.g.,][]{2008ApJ...681.1484R}, in contrast to the large disk
fractions of pre-main sequence (pre-MS) star populations associated
with molecular clouds \citep[e.g.,][]{2008ARA&A..46..339W}. This
contrast reflects the relatively advanced ages of the known young
stars within $\sim100$ pc and suggests that, while many may
still be building terrestrial planets \citep{2010ApJ...717L..57M},
most have already moved beyond the epoch of active giant planet
formation.

Examples of {\it actively accreting}, low-mass pre-MS stars (i.e.,
classical T Tauri stars; cTTS) within $\sim$100 pc are rarer still
\citep{2008hsf2.book..757T}.  The archetype of these nearby cTTS
systems --- which are characterized by their unusually large H$\alpha$
emission equivalent widths --- is the intensively studied TW Hya. At a
distance of 56 pc, TW Hya is among the closest known cTTS\footnote{The
  TW Hya Association (TWA) members Hen 3-600 \cite[$D \sim 45$
  pc;][and references therein]{2007ApJ...671..592H} and TWA 30AB
  \citep[$D = 42$
  pc;][]{2010arXiv1008.4340L,2010ApJ...714...45L} are the closest
  known examples of actively accreting, low-mass stars.}. With an
estimated age of $\sim8$ Myr
\citep{2004ARA&A..42..685Z,2008hsf2.book..757T}, TW Hya also remains one of
the oldest known examples of a cTTS. In the latter regard, however, it
is rivaled by two other, nearby systems: the close binary cTTS V4046
Sgr \citep[age $\sim12$ Myr, distance 72 pc;][and references
therein]{2008hsf2.book..757T,2008A&A...492..469K} and the cTTS MP Mus
\citep[= PDS
66;][]{2002AJ....124.1670M,2007A&A...465L...5A,2009ApJ...697.1305C}.
Estimates of the age and distance of the latter star range,
respectively, from 6 Myr to 17 Myr and from 86 pc to 103 pc
\citep{2002AJ....124.1670M,2008hsf2.book..757T}.  The younger age
(hence larger distance) appears more accurate, given the Li absorption
line strength of MP Mus \citep{2010arXiv1005.0984W} and its likely
association with the $\epsilon$ Cha group \citep{2008hsf2.book..757T}.
High-resolution (gratings) X-ray spectra of TW Hya, V4046 Sgr, and MP
Mus reveal evidence for a significant emission contribution from
accretion shocks, as opposed to coronal activity
\citep{2002ApJ...567..434K,2004A&A...418..687S,2006A&A...459L..29G,2007A&A...465L...5A}. This
is consistent with observations demonstrating that the H$\alpha$ and
UV emission from each of these three stars is stronger than that
typically associated with pure chromospheric activity
\citep{2007ApJ...671..592H}.

Radio molecular line emission studies of circumstellar planet-forming
disks provide unique insight into the Jovian planet, Kuiper Belt, and
comet formation zones within the outer regions (tens to hundreds of AU
in radius) of pre-MS circumstellar disks
\citep[e.g.,][]{1995Natur.373..494Z,1997A&A...317L..55D,2004A&A...425..955T}.
Thus far, however, only four star-disk systems within $\sim100$ pc of
Earth have been detected in molecular emission lines with a radio
telescope: the aforementioned $\sim10$ Myr-old, K-type cTTS systems TW
Hya \citep{1995Natur.373..494Z,1997Sci...277...67K} and V4046 Sgr
\citep{2008A&A...492..469K}, and the A-type stars 49 Cet and HD 141569
\citep[ages $\sim20$ and $\sim5$ Myr,
respectively;][]{1995Natur.373..494Z}. Detections of a suite of
molecular transitions toward TW Hya established that it possesses a
rich molecular disk viewed nearly pole-on
\citep{1997Sci...277...67K,2004A&A...425..955T}, while the
double-peaked molecular line profiles characteristic of Keplerian
rotation observed toward V4046 Sgr \citep{2008A&A...492..469K}
similarly demonstrated that this close (period 2.4 d) binary system is
orbited by a circumbinary molecular disk. These single-dish radio
molecular line spectroscopy results have subsequently been confirmed
via radio interferometric (Submillimeter Array) imaging of CO emission
\citep{2004ApJ...616L..11Q,2006ApJ...636L.157Q,2010ApJ...720.1684R}.

    In an effort to similarly characterize the gas mass, dynamics, and
    chemistry within the disk of MP Mus (spectral type K1 IVe), we
    searched for, and
    detected, CO emission from MP Mus with the Atacama Pathfinder
    Experiment (APEX) 12 m telescope\footnote{APEX
  is a collaboration between the Max-Planck-Institut fur
  Radioastronomie, the European Southern Observatory, and the Onsala
  Space Observatory.}. In this paper, we report on
    these results and their interpretation. We also
    report the nondetection of CO emission from a sample of ten other
    nearby ($D\stackrel{<}{\sim}100$ pc), dusty,
    young (ages $\sim10$--100 Myr) stars of spectral type A--G with
    the Institut de Radio
    Astronomie Millimetrique (IRAM) 30 m telescope.

\section{Observations and Results}

\subsection{IRAM 30 m telescope CO observations of nearby, 
dusty young stars}

We observed the ten stars\footnote{During the 2009 August IRAM 30 m
  observing run we also observed the young late-A/early-F star
  HR 8799, which is orbited by multiple giant planets and debris disks
  \citep{2008Sci...322.1348M,2009ApJ...705..314S}. As mentioned by
  \citet{2009ApJ...705..314S}, our $^{12}$CO(2--1) spectrum
  established that bright interstellar emission is present at the
  radial velocity of the star, precluding any attempt to constrain its
  circumstellar gas mass via single-dish CO measurements.}  in
Table~\ref{tbl:nondetections} in the 230.538 GHz $J=2\rightarrow 1$
transition of $^{12}$CO with the IRAM 30 m telescope in August
2009. These stars (Table~\ref{tbl:nondetections}) were chosen from the
list of \citet{2007ApJ...660.1556R} on the basis of large infrared
excesses (indicative of disk dust masses $M_d\stackrel{>}{\sim} 0.05
M_\earth$) and estimated ages $\stackrel{<}{\sim}$ 100 Myr. We
observed all target stars using IRAM's EMIR receivers and VESPA
autocorrelator, with velocity resolution of 0.8 km s$^{-1}$. Spectra
were acquired in wobbler switching mode, resulting in flat baselines
for individual spectra. Total on-source integration times were between
1 and 2 hours per star. The weather was generally good ($\tau_{225}
\sim$ 0.25--0.4) throughout the observations, with time-averaged
system temperatures in the range 220--510 K. Pointing and focus were
checked (against standard pointing sources and planets) every $\sim$2
hours. Typical pointing errors were $\sim3''$, i.e., $\sim$1/4
beamwidth (FWHP $12''$ at 230 GHz).

We used the CLASS\footnote{See http://iram.fr/IRAMFR/GILDAS/} radio
spectral line data reduction package to sum over individual spectral
scans obtained for each star, and then to subtract a linear-fit
baseline from each of the resulting integrated spectra, calculating
channel-to-channel noise levels in the process. No CO emission lines
were evident in any of the resulting spectra obtained for the ten
sample stars. Upper limits on main-beam brightness temperature
$T_{mb}$ (assuming $F_{eff}=0.94$ and $\eta_{mb}=0.52$ for
measurements at 230 GHz with the IRAM 30 m telescope) and
velocity-integrated CO line intensity $I$ are reported in
Table~\ref{tbl:nondetections} as the values $3\sigma_T$ and
$3\sigma_I$, respectively, where $\sigma_T$ is the
efficiency-corrected rms channel-to-channel noise in the CO spectrum
and $\sigma_I = \sigma_T \sqrt{\Delta v \delta v}$, with velocity
resolution $\delta v = 0.8$ km s$^{-1}$ and an assumed linewidth
$\Delta v = 3$ km s$^{-1}$.

\subsection{APEX 12 m telescope CO observations of MP Mus}

Service-mode APEX 12 m telescope observations of MP Mus (J2000
coordinates $\alpha=$13:22:07.55, $\delta=$$-$69:38:12.2) in the 345.796
GHz $J=3\rightarrow 2$ transition of $^{12}$CO were performed on 15
and 16 April 2010 UT. The receiver was SHFI/APEX2, using the facility
FFTS backend with velocity resolution 0.21 km s$^{-1}$. Spectra were
acquired in position switching mode using the wobbler. Total on-source
integration time was 138 min. The weather was generally good
throughout the observations ($\tau_{225} \sim$ 0.2; pwv 1.8 mm;
time-averaged system temperature $\sim500$ K). The best focus was
established at the beginning of each day's observations, and regular
pointing checks (using nearby planets and quasars as references)
indicated pointing errors were $\stackrel{<}{\sim} 2''$, i.e.,
$\sim\frac{1}{8}$ beamwidth (FWHP $17.5''$ at 345 GHz). APEX spectra
of MP Mus were reduced using CLASS, as described in \S 2.1.

The resulting, integrated, baseline- and beam-efficiency-corrected
spectrum (assuming $\eta_{mb} = 0.73$ for measurements at 345 GHz
with the APEX 12 m telescope) is presented in
Fig.~\ref{fig:CO32profcomp}. This spectrum clearly reveals the
detection of $^{12}$CO(3--2) emission from MP Mus. From a Gaussian fit
to the line profile, we find the central velocity to be $+3.9
\pm 0.3$ km s$^{-1}$ with respect to the local standard of rest
(LSR). This translates to a heliocentric velocity of $11.3 \pm
0.3$ km s$^{-1}$, which is consistent with the systemic velocity of MP
Mus as determined from optical spectroscopy \citep[$11.6\pm 0.2$ km
s$^{-1}$;][]{2006A&A...460..695T}. The integrated line intensity obtained
from the best-fit Gaussian is $0.21 \pm 0.03$ K km s$^{-1}$.  

\section{Analysis}

\subsection{MP Mus: constraints on molecular disk radius and mass}

\begin{figure}[htb]
  \centering
\includegraphics[width=6in,angle=0]{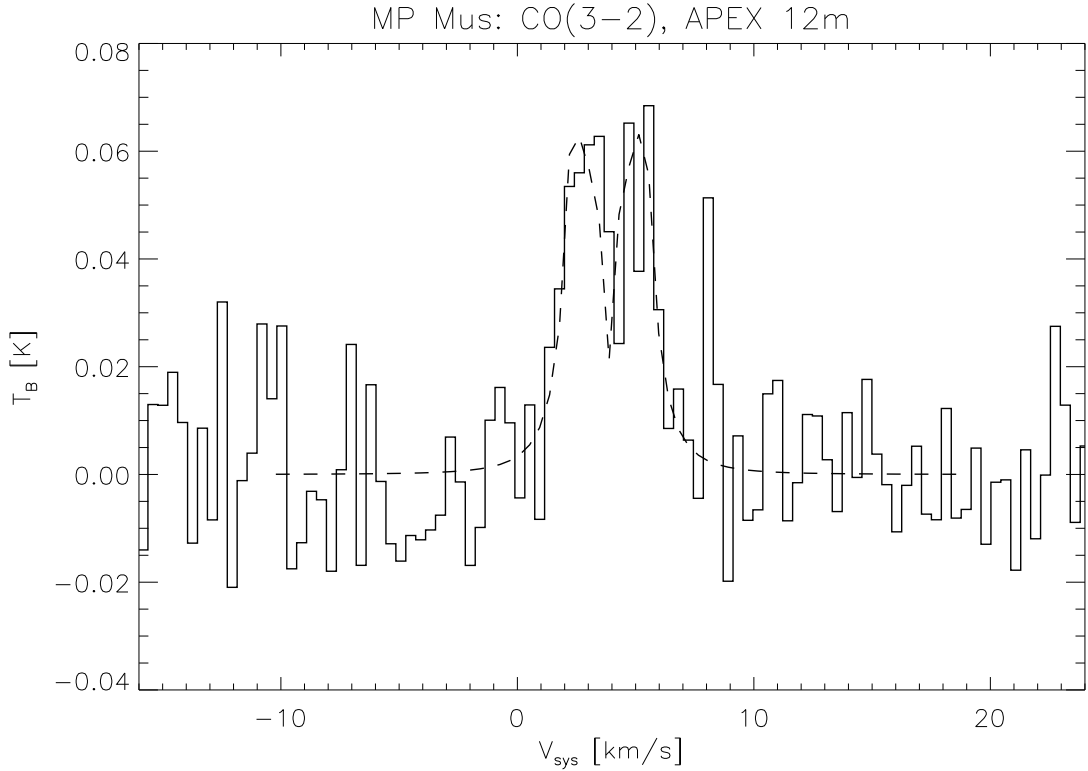}
\caption{$^{12}$CO(3--2) line profile of MP Mus as measured by the
  APEX 12 m telescope and spectrometer (solid line) overlaid with
  best-fit Keplerian disk model profile (dashed line). Ordinate is
  velocity with respect to the local standard of rest and abscissa is
  main-beam brightness temperature. The spectrometer data have been
  Hanning smoothed to a velocity resolution of $0.4$ km s$^{-1}$. The
  rms noise level of the spectrum is 13 mK per $0.4$ km s$^{-1}$
  bin.}
\label{fig:CO32profcomp} 
\end{figure}

Given the signal-to-noise ratio of the CO(3--2) data obtained for MP
Mus (Fig.~\ref{fig:CO32profcomp}), we cannot unambiguously determine
the underlying emission line profile. However, the steep sides and
possible central valley (i.e., double-peaked appearance) of the MP Mus
CO emission line are features expected in the case of an orbiting
molecular disk \citep[e.g.,][]{1993ApJ...402..280B}. Hence, we fit the
profile with a parametric representation of the line profile predicted
by the Keplerian disk model of \citet{1993ApJ...402..280B} as described in
\citet{2008A&A...486..239K}. Provided the observed line profile is
well fit, this method yields an estimate of outer disk rotation
velocity $v_d$ --- where $v_d$ is equivalent to the half-value of the
velocity separation of the red and blue peaks in the double-peaked
line profile predicted for Keplerian rotation --- as well as measures
of the slopes of the inner and outer portions of the line profiles
($p_d$ and $q$, respectively). The value of $q$ serves as an
indication of the slope of the disk radial temperature profile ($T
\propto r^{-q}$), while $p_d$ indicates the degree of central filling
of the line profile by emission from gas at low radial velocities
(e.g., for a nearly edge-on disk, $p_d=1$ would correspond to a sharp
outer edge and values $p_d<1$ would indicate lack of a sharp edge).

The observed line profile is overlaid with the best-fit model profile
in Fig.~\ref{fig:CO32profcomp}. The comparison confirms that the CO
line emission profile of MP Mus is consistent with that of a molecular
disk in Keplerian rotation. The best-fit Keplerian model parameter
values (and formal uncertainties on these values) are $v_d = 1.6 \pm
0.16$ km s$^{-1}$, $q = 0.5 \pm 0.2$, and $p_d = 0.25 \pm 0.1$. The
best-fit value of $q$ is compatible with standard models of Keplerian
disks \citep[e.g.,][]{1993ApJ...402..280B}.

Given an estimate for the mass of MP Mus, $M_\star$, the best-fit
value of $v_d$ can be used to place constraints on the molecular disk
outer cutoff radius $R_{out,CO}$ for an assumed disk inclination $i$
\citep[e.g.,][and references therein]{2008ApJ...683.1085Z}. Adopting
$M_\star = 1.2$ $M_\odot$ \citep[based on comparison with pre-MS
evolutionary tracks;][]{2002AJ....124.1670M} and $i = 32^\circ$
\citep[based on the morphology of the disk as detected in
coronagraphic imaging of starlight reflected off of
dust;][]{2009ApJ...697.1305C}, the result $v_d = 1.6$ km s$^{-1}$
implies $R_{out,CO} \approx 120$ AU. This is somewhat smaller than the
outer disk radius inferred (also on the basis of coronagraphic
imaging) by \citet{2009ApJ...697.1305C}, i.e., $R_{out,d} \approx 170$
AU, assuming a distance to MP Mus of 86 pc.  However, given the
uncertainties in our line profile analysis as well as in the mass,
age, and distance of MP Mus (\S 4) --- and the possibility that the
(3--2) transition of CO is not well excited within the outermost
regions of the disk --- it appears our estimate of the radius of the
molecular disk is in reasonable agreement with the dust disk radius as
measured coronagraphically.

The best-fit Keplerian model $^{12}$CO(3--2) line profile area is
slightly larger than that obtained from the Gaussian fit (\S 2.1),
i.e., $0.26 \pm 0.03$ K km s$^{-1}$. This is a factor $\sim7$ weaker
than the integrated $^{12}$CO(3--2) line intensity of TW Hya as
measured with the JCMT \citep{1997Sci...277...67K}, suggesting that
the intrinsic total $^{12}$CO(3--2) line intensity of MP Mus is a
factor $\sim3$ smaller than that of TW Hya, after accounting for the
difference in JCMT and APEX beam sizes and assuming MP Mus lies at
$\sim 100$ pc (\S 4). The measured total APEX $^{12}$CO(3--2) line
intensity of MP Mus is equivalent\footnote{See
  http://www.apex-telescope.org/telescope/efficiency/} to an
integrated line flux of $10.5 \pm 1.5$ Jy km s$^{-1}$. This result can
be used to estimate the molecular gas mass of the MP Mus disk
\citep[e.g.,][and references therein]{2008ApJ...683.1085Z} --- albeit
with large uncertainties, especially given that thus far we have no
measurements of other CO isotopes and (hence) no means to estimate the
$^{12}$CO(3--2) line optical depth ($\tau_{32}$) other than the
foregoing rough comparison with TW Hya. Adopting the standard
assumptions of a CO:H$_2$ number ratio of $10^{-4}$ and molecular gas
excitation temperature $T_{ex}=40$ K, and assuming $\tau_{32}=3$
\citep[i.e., a factor $\sim3$ smaller than that inferred for TW
Hya;][and references therein]{2008A&A...492..469K} we obtain a
molecular (H$_2$) gas mass of $\sim9\times10^{-6}$ $M_\odot$ (i.e.,
$\sim3 M_\earth$). This inferred disk gas mass, which scales
approximately linearly with $\tau_{32}$ if the $^{12}$CO(3--2) is
optically thick, is somewhat less than the circumstellar dust mass of
17 $M_\earth$ determined for MP Mus by \citet{2005AJ....129.1049C} on
the basis of the star's 1.2 mm flux.

\subsection{Other nearby, dusty, young stars: gas mass upper limits}

We used the same method just described for MP Mus to estimate
molecular gas mass upper limits for the dusty disks of the ten stars
observed with, but not detected by, the IRAM 30 m. The results,
obtained under the assumptions CO:H$_2$ $=10^{-4}$, $T_{ex}=40$ K, and
$\tau_{21}=1$, are reported in Table~\ref{tbl:nondetections}. These
upper limits are of course subject to the same uncertainties just
described in the case of MP Mus. Nevertheless, the results in
Table~\ref{tbl:nondetections} indicate that, for these ten stars, the
disk gas masses are constrained to be of order of (or smaller than) a few
times the dust masses inferred from far-IR (IRAS) data, provided the
CO emission is not optically thick or that gas-phase CO in these disks
is not severely depleted relative to H$_2$.

\section{Discussion}

With our CO detection of a disk orbiting MP Mus, this star joins
$\sim$8 Myr-old TW Hya and $\sim12$ Myr-old V4046 Sgr as the only
late-type (therefore low-mass) pre-MS stars within $\sim100$ pc of
Earth that are known to retain orbiting molecular disks. In all three
cases, the circumstellar gas masses inferred from far-IR and radio
emission lines --- a few to a few hundred $M_\earth$ \citep[][and
references
therein]{1997Sci...277...67K,2010ApJ...720.1684R,2010A&A...518L.125T}
--- are similar to the dust masses determined from far-IR and (sub)mm
photometry. This implies dust-to-gas ratios near unity in the outer
($\stackrel{>}{\sim}10$ AU) regions of these disks and/or, in the
case of gas mass estimates based on radio CO lines, that
gas-phase CO is severely depleted relative to H$_2$.

The overall similarity of MP Mus to the TW Hya and V4046 Sgr star-disk
systems favors the younger end of the range of ages estimated to date
for MP Mus. This is consistent with the recent results of
\citet{2010arXiv1005.0984W}, who estimated an age of $7 \pm 3$ Myr
based on the Li absorption line strength of MP Mus. An age of $\sim7$
Myr would, in turn, imply the distance to MP Mus is $\sim100$ pc
\citep{2008hsf2.book..757T}. For such a distance, the mass of MP Mus
could be $\sim20$\% larger than previously determined (i.e., closer to
1.4 $M_\odot$), depending on the adopted theoretical pre-MS
evolutionary tracks \citep[][and references
therein]{2002AJ....124.1670M}. If so, the CO and dust disk radii would
be correspondingly larger than the estimates stated in \S 3.1.

Given the relatively advanced pre-MS evolutionary states of TW Hya,
V4046 Sgr, and MP Mus, their circumstellar environments likely are in
transition from predominantly primordial material to ``debris'' disks
that are populated by dust grains generated via collisions between
rapidly growing planetesimals \citep[][and references
therein]{2009ApJ...697.1305C}. The stars for which we failed to detect
molecular gas (Table~\ref{tbl:nondetections}) likely fall into this
latter category; all of the Table~\ref{tbl:nondetections} stars have
disk dust masses much smaller than those of TW Hya, V4046 Sgr, and MP
Mus, and all have ages $\ge$10 Myr. The comparison of the
residual gas masses of the three nearby cTTS with
the gas mass upper limits in Table~\ref{tbl:nondetections} thereby
further reinforces the notion that the gaseous disks required to spawn
Jovian planets dissipate rapidly, i.e., within the first few Myr of
pre-MS evolution \citep{1995Natur.373..494Z}.

``Transition disks'' usually display evidence for inner (radii $\sim$
a few AU) disk clearings that may be the results of recent or ongoing
planet formation \citep[e.g.,][]{2005ApJ...630L.185C}. In this
respect, MP Mus, TW Hya, and V4046 Sgr are typical of transition disk
systems; all three lack the significant near-IR excesses, indicative
of hot dust in inner disks, that are characteristic of rapidly
accreting cTTS in molecular clouds
\citep[e.g.,][]{1997AJ....114..288M}. At the same time, like certain
cTTS in Taurus that display transition disks \citep[DM Aur and GM
Aur;][]{2005ApJ...630L.185C}, all three nearby, late-type stars with
detectable circumstellar CO emission also display spectral evidence
for ongoing accretion --- evidence that (to our knowledge) is lacking
in the cases of the Table~\ref{tbl:nondetections} stars. The fact that
TW Hya, V4046 Sgr and MP Mus are evidently still accreting from their
``transition disks'' may indicate that they are orbited by young
planets \citep[see discussion in][]{2007ApJ...655L.105S}.
Regardless, the detections of CO
in the dusty disks of MP Mus, TW Hya and V4046 Sgr suggest that the
presence of residual disk molecular gas is closely linked to (likely
enables) sustained accretion activity at advanced pre-MS ages.


\acknowledgments{This research was supported via NASA
  Astrophysics Data Analysis grant NNX09AC96G to RIT and UCLA. The
  authors wish to thank the staff of the APEX telescope for their
  expertise in carrying out the observations of MP Mus.}


\newpage

\begin{sidewaystable}
\caption{IRAM 30 m CO Observations of Nearby, Dusty Stars}
\label{tbl:nondetections}
\begin{center}
\begin{tabular}{lcccccccccc}
  \hline
  HD & $\alpha$ & $\delta$ & sp.\ type$^a$ & $D^a$   & age$^a$   & $M_d$$^a$ & $3\sigma_T$ 
  & \multicolumn{2}{c}{$3\sigma_I$}  & $M_g$ \\
  & \multicolumn{2}{c}{(J2000)} &           & (pc)  & (Myr) & ($M_\earth$) & (mK) 
  & (K km s$^{-1}$) & (Jy km s$^{-1}$) &  ($M_\earth$) \\
  \hline
  \hline
  15745 & 02:32:55.81 & 37:20:01.0 & F0 & 64 & 30? & 0.09 & 75 & 0.40 & 3.2 & $<0.6$ \\
  21997 & 03:31:53.64 & $-$25:36:50.9 & A3 & 74 & 50? & 0.22 & 126 & 0.66 & 5.2 & $<1.3$ \\
  30447 & 04:46:49.52 & $-$26:18:08.8 & F3 & 78 & $<100$ & 0.13 & 147 & 0.77 & 6.0 & $<1.7$ \\
  32297 & 05:02:27.43 & 07:27:39.6 & A0 & 112 & 20? & 0.46 & 93 & 0.50 & 3.9 & $<2.2$ \\
  38206 & 05:43:21.67 & $-$18:33:26.9 & A0 & 69 & 50 & 0.06 & 156 & 0.81 & 6.4 & $<1.4$\\
  85672 & 09:53:59.15 & 27:41:43.6 & A0 & 93 & 30? & ... & 69 & 0.36 & 2.8 & $<1.1$ \\
  107146 & 12:19:06.50 & 16:32:53.8 & G2 & 29 & $<100$ & 0.09 & 57 & 0.30 & 2.4 & $<0.09$\\
  131835 & 14:56:54.46 & $-$35:41:43.6 & A2 & 111 & 10 & 0.2 & 207 & 1.1 & 8.4 & $<4.8$ \\
  191089 & 20:09:05.21 & $-$26:13:26.5 & F5 & 54 & $<30$ & 0.03 & 132 & 0.69 & 5.4 & $<0.7$ \\
  221853 & 23:35:36.15  &  08:22:57.4 & F0 & 71 & $<100$ & 0.05 & 57 & 0.30 & 2.4 & $<0.6$ \\
  \hline
\end{tabular}
\end{center}

\footnotesize
a) Spectral types, distances, ages, and estimated dust masses as listed in 
\citet{2007ApJ...660.1556R}.
\end{sidewaystable}

\newpage

\end{document}